\documentclass{icrc2009}

\usepackage{graphicx}   % for including figures
\usepackage{caption}    % for captions
\usepackage[font=footnotesize]{subfig} % subfig.sty for a double column floating figure using two subfigures
\usepackage{fixltx2e}
\usepackage{url}

\newcommand{\shorttitle}[1]%
{\markboth{Proceedings of the 31\MakeLowercase{$^{st}$} ICRC, {\L}\'{o}d\'{z} 2009}{#1} }
\newcommand{\etal}{\MakeLowercase{\textit{et al. }}} % "et al."

\begin{document}
\title{Influence of solar and cosmic-ray variablity on climate}
\author{\IEEEauthorblockN{Badruddin\IEEEauthorrefmark{1}, O.P.M. Aslam \IEEEauthorrefmark{1} and
			  M. Singh\IEEEauthorrefmark{2}}
                            \\
\IEEEauthorblockA{\IEEEauthorrefmark{1}Department of Physics, A.M.U., Aligarh, India - 202002}
\IEEEauthorblockA{\IEEEauthorrefmark{2}Department of Physics, H.I.T., Greater Noida, India - 201306}}

% please write the preseter's name and short title (3-4 words maximum)
%    which will appear at the header of the even pages.
\shorttitle{Badruddin \etal Solar and CR variablity on climate}
\maketitle

\begin{abstract}
 We analyze solar, geomagnetic and cosmic ray flux data along with rainfall and temperature data for almost five solar cycles. We provide evidence of significant influence of solar variability on climate. Specifically, we demonstrate association between lower (higher) rainfall and higher (lower) temperatures with increasing (decreasing) solar activity and decreasing (increasing) cosmic ray intensities. We propose a plausible scenario that accounts the results of our analysis.\\
   \end{abstract}

 
\section{Introduction}
 The relationship between solar variability and terrestrial weather/climate is an interesting and controversial subject. Past studies of possible relationship between solar activity and temperature have been found to be positive, negative and even no correlation between them. Such contradictory results still elude the scientific community upon complete understanding of the influence of the sun on weather/climate of our planet [1,2].\\
 Both the Indian rainfall and temperature are, in general, inversely related to sunspot number, although the results of some of the studies are conflicting (e. g. see [3-5] and references their in).\\
The influence of solar activity on the Indian monsoon rainfall has been studied recently by [6-8] using Indian rainfall data. found that spring and southwest monsoon rainfall variability has significant positive correlations with sunspot activity during the corresponding period [7].  A found that the average rainfall is higher during periods of greater solar activity [6]. Earlier studies related to association between the sunspot numbers and monsoon rainfall variability show a moderate to strong correlation. It was suggested that one of the primary controls on centennial to decadal scale changes in tropical rainfall and monsoon intensity is variation in solar activity [9].\\
GCR-Cloud Cover-Climate relationship has been reported to be significant and controversial too (e.g. see [10-21]); see an excellent review on the topic by [22] and references therein). Thus the whole area of sun-climate relationship is complex and needs further study.\\
 
\section{Results and discussion}
 We have utilized monthly average data of All India rainfall and maximum temperature (representing Indian weather/climate), solar flare index and geomagnetic activity index (representing solar/geomagnetic variability) and galactic cosmic ray flux for about five solar cycles (1953-2005). We have divided total 53 years from 1953 to 2005 into five groups on the basis of amount of All India Summer Monsoon Rainfall (ASMR) during the summer monsoon months of June, July, August and September. One-fifth (20\%) of total 53 years with lowest ASMR (LSMR) and same number of years (20\%) with highest ASMR (HSMR) were considered for the purpose of analysis. Average behavior of variations in various parameters during different monsoon months (June, July, August and September) were obtained by analyzing the solar, geomagnetic, cosmic ray and climate data, by applying the method of superposed epoch analysis. Using the same procedure, the average behavior of these parameters was obtained for each month for total period (1953-2005). The values so obtained for each month were then subtracted from group values (LSMR, HSMR) of respective months. In this way deviations in certain parameters e.g. Rainfall (mm), Tmax ($^{0}$C), were obtained for each month in both LSMR and HSMR groups. The results so obtained are plotted in Figs. 1-5.\\
From these Figures we see that decreasing solar activity during summer monsoon months, as evident from vitiation in solar flare index and AA index (Fig. 1 \& 2) and increasing cosmic ray intensity (Fig. 3) is associated with high rainfall (Fig. 4) and lower temperature (Fig. 5). We also see from these figures that increasing solar activity during summer monsoon months and decreasing cosmic ray intensity is associated with low rainfall and higher temperature (see Figs. 1-5).\\
Some of the previous studies have suggested that on decadal, multi-decadal, and centennial scale, variability in solar activity might have significant impact on regional climate and could have caused severe draughts and floods in the past (e.g. [23] and references therein). Studies of 6000-years record of changes in drought and precipitation in northern China has led [24] to conclude that the wide spread global drought variability is consistent with the assumption of an external global force such as solar force. They found that most of the dry and warm periods over the last 6000-year correspond well with stronger solar activity and relativity wet and cold periods correspond well with relatively weaker solar activity.\\
Periodicities in occurrence of rainfall variability are almost similar to periodicities in the sunspot occurrence activity [7]. Another study [25] of rainfall and drought patterns in equatorial east Africa during the past 1100-year reported severe drought during periods coinciding with phases of high solar activity with intervening epochs of increased precipitation periods of low solar activity. These results contrasts that of [23] who found that, on decadal and multi-decadal time scale, the intensity of Indian monsoon have decreased during periods of solar minimum during last millennium. Most of earlier studies have observed solar influence on climate on very long timescale (decadal, multi-decadal, centennial etc.). In this paper we demonstrated solar/cosmic ray variability effects on monthly scale, probably for the first time.\\

% see \section{Examples of \LaTeX\  instructions} and \subsection{Figures}
 \begin{figure}[!t]

  \centering
  \includegraphics[width=2.7in]{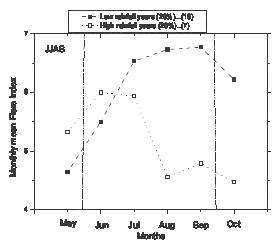}
  \caption{Average value of solar flux index from May to October during HSMR and LSMR years.}
   \end{figure}
    \begin{figure}[!t]
  \centering
  \includegraphics[width=2.7in]{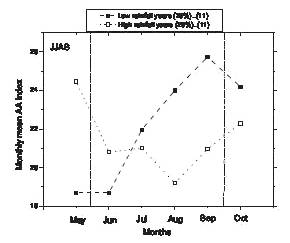}
  \caption{Average value of AA index from May to October during HSMR and LSMR.}
   \end{figure}
     \begin{figure}[!t]
  \centering
  \includegraphics[width=2.7in]{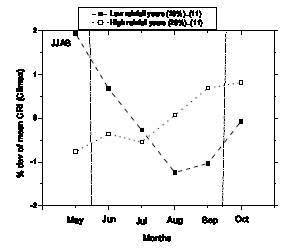}
  \caption{Percentage deviation of average cosmic ray intensity from May to October during HSMR and LSMR.}
   \end{figure}
   \begin{figure}[!t]
  \centering
  \includegraphics[width=2.7in]{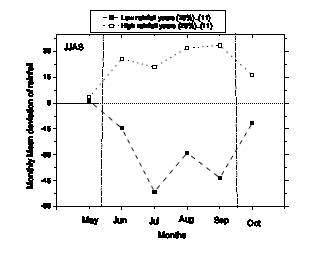}
  \caption{Average deviation of rainfall from all 53 years values during HSMR AND LSMR for months May to October.}
   \end{figure}
    \begin{figure}[!t]
  \centering
  \includegraphics[width=2.7in]{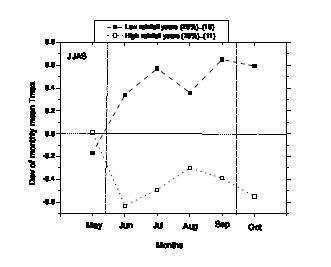}
  \caption{Average deviation of temperature from all 53-year values in respective months during HSMR and LSMR for months May to October.}
   \end{figure}

% see \section{Examples of \LaTeX\  instructions}  \subsection{Tables}
  \begin{table}[!h]
  \caption{Change in parameters from May to October, Including summer monsoon months of June, July, August and September.}
  \label{table_simple}
  \centering
  \begin{tabular}{|c|c|c|}
  \hline
   Parameters  &  Low rainfall & High rainfall \\
   \hline 
    Flare index    &   2.13  ($\uparrow$)  &  1.53  ($\downarrow$)\\
    AA Index    &   7.04  ($\uparrow$) &   5.28  ($\downarrow$) \\
    Dev\(CRI\)   &   3.17   ($\downarrow$) &   1.58  ($\uparrow$) \\
    Dev of rainfall(mm)    &   52.71   ($\downarrow$) &   30.03  ($\uparrow$)\\
    Dev of Tmax($^{0}$C)    &   +0.82 ($\uparrow$)  &   -0.64  ($\downarrow$)\\
  \hline
  \end{tabular}
  \end{table}

% see \section{Examples of \LaTeX\  instructions} and \subsection{Figures}
% An example of a double column floating figure using two subfigures.
% The double column figure must be placed in the source text file 
%         within the text of the previous page

   %\label{double_fig}
 %\end{figure*}

 \noindent
 \noindent

\section{conclusions}
   Our results suggest the influence of solar and/or cosmic ray variability on Indian summer monsoon rainfall and Indian climate. Decreasing solar activity and/or increasing cosmic ray intensity are associated with higher rainfall and lower temperature. These results provide evidence for the role of solar variability on climate variability possibly through cosmic rays, probably for the fast time at seasonal and even monthly time scale.\\

% see \section{Examples of \LaTeX\  instructions} and \subsection{Figures}
% An example of a double column floating figure using two subfigures.
% The double column figure must be placed in the source text file 
%         within the text of the previous page


\begin{thebibliography}{99}
 \bibitem{hoyt}   D.V.~Hoyt and K.H.~Schatten, {\it Oxford: Oxford University Press.,} 125-142, 1997.
   \bibitem{kane}     R.P.~Kane, {\it Indian Space Research Organisation, Bangalore, Scientific Note.,} ISRO-SN-11-99, 1999.
   \bibitem{bhal}     H.N.~Bhale, R. S.~Reddy, D. A.~Mooley and B. V.~Ramanamurty, {\it Earth and Planet. Sci. Lett.,} 90, 245-262, 1981.
   \bibitem{jagg}     P.~Jagannathan and H.N.~Bhalme, {\it Mon. Weather Rev.,} 101, 681-700, 1973.
   \bibitem{part}     B.~Parthasarthy and D.A.~Mooley, {\it Mon. Weather Rev.,} 106, 771-781, 1978.
   \bibitem{bhat}     S.~Bhattacharya and R.~Narasimha, {\it Geophys. Res. Lett.,} 32, L05813, 2005.
   \bibitem{hire}     K.M.~Hiremath and P.I.~Mandi, {\it New Astron.,} 9, 651-652, 2004.
   \bibitem{badr}     Badruddin, Y.P.~Singh  and M.~Singh, {\it Proceedings of the ILWS Workshop,}  444-447, 2006.
  \bibitem{neff}     U.~Neff,  S.J.~Burns, A.~Mangini,  M.~Mudelsee,  D.~Fleitmann and A.~Matter, {\it Nature.,} 411, 290-293, 2001.
    \bibitem{stoz}   Stozhkov,  Yu. I. et al., {\it Il Nuovo Cim.,} 18c, 335-341, 1995.
   \bibitem{cars}     K.S.~Carslaw, R.G.~Harrison and J.~Kirkby, {\it Science.,} 298, 1732-1737, 2002.
   \bibitem{erly}     A.D.~Erlykin, G.~Gayali, K.~Kudela,  T.~Solan and A.W.~Wolfendale, {\it JASTP (accepted).,} 2009.
   \bibitem{harr}     R.J.~Harrison and D.B.~Stphenson, {\it Proc. R. Soc.A.,} 462, 1221-1233, 2006.
   \bibitem{jorg}     T.B.~Jorgenson and A.W.~Hansen, {\it J. Atmos. Sol. Terr. Phys.,} 62, 73-77, 2000.
   \bibitem{kant}     S.C.~kanthaler,  R.~Toumi and J.D.~Haigh, {\it Geophys. Res. Lett.,} 26, 863-865, 1999.
   \bibitem{kris}     J.E.~Hristjansson, J.~Kristiansen, {\it Geophys. Res. Lett.,} 105, 11851-11863, 2000.
   \bibitem{mars}     N. D.~Marsh and H.~Sevnsmark, {\it Phys. Rev. Lett.,} 85(23), 5004-5007, 2000.
   \bibitem{pudo}     M.I.~Pudovkin and S.V.~Veretenko, {\it J. Atmos. Sol. Terr.Phys.,} 59, 1739-1746, 1997.
   \bibitem{sven}     H.~Svensmark and E.~Friss-Christensen, {\it J. Atmos. Sol. Terr. Phys.,} 59, 1225, 1997.
   \bibitem{usos}     I.G.~Usoskin, N.D.~Marsh,  G.A.~Kovaltsov, K.~Mursula  and O.G.~Gladysheva, {\it Geophys. Res. Lett.,} 31, L16109, 2004.
   \bibitem{Sloa}     A.W.~Solan and Wolfendale, {\it Environ. Res. Lett.,} 3, 024001, 2008.
  \bibitem{kirk}     J.~Kirkby  and Surv, {\it Geophys. Res. Lett.,} 28, 333-375, 2007.
\bibitem{agni}     R.~Agnihotri, K.~Dutta, R.~Bhusan and B. L. K.~Somayajulu, {\it Earth and Planet. Sci. Lett.,} 198, 521-527, 2002.
   \bibitem{hong}     Y.T.~Hong,  Z.G.~Wang,  H.B.~Jiang,  Q.H.~Lin,  B.~Hong,  Y.Z.~Zhu,  Y.~Wang, L.S.~Xu,  X.T.~Leng, and H.D.~Li, {\it Earth and Planet. Sci. Lett.,} 185, 111-119, 2001.
   \bibitem{vers}     D.~Verschuren,  K.R.~Laird,  and B. F.~Cumming, {\it Nature.,} 403, 410-414, 2000.

\end{thebibliography}
\end{document}